\documentclass[10pt,aps,prd,superscriptaddress,nofootinbib,nobibnotes,longbibliography,floatfix,twocolumn]{revtex4-2}

\usepackage{bm}
\usepackage{mathtools,
amsmath,
amssymb,
amsfonts,
mathrsfs,
chngcntr,
multirow}

\let\cc\corresponds
\let\corresponds\relax
\usepackage{mathabx}
\let\corresponds\cc

\usepackage{booktabs}
\usepackage[utf8]{inputenc}
\usepackage[T1]{fontenc}
\usepackage[dvipsnames]{xcolor}
\usepackage[unicode]{hyperref}
\hypersetup{colorlinks=true, citecolor=MidnightBlue,
            linkcolor=MidnightBlue, urlcolor=MidnightBlue, linktocpage=true}
\usepackage[normalem]{ulem}
\usepackage{orcidlink}
\usepackage[capitalize]{cleveref}

\newcommand{\orcid}[1]{\href{https://orcid.org/#1}{\includegraphics[width=10pt]{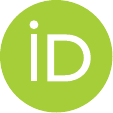}}}

\begin{document}
\title{Constraining deviations from the Teukolsky equation with GW250114}

\author{Sebastian H.\,V\"olkel\,\orcid{0000-0002-9432-7690}}
\email{sebastian.voelkel@uni-tuebingen.de}
\affiliation{Theoretical Astrophysics, IAAT, University of T\"ubingen, Auf der Morgenstelle 10, D-72076 T\"ubingen, Germany}
\affiliation{Max Planck Institute for Gravitational Physics (Albert Einstein Institute), \\Am Mühlenberg 1, Potsdam 14476, Germany}

\author{Nicola Franchini \orcid{0000-0002-9939-733X}}
\email{nicola.franchini@tecnico.ulisboa.pt}
\affiliation{CENTRA, Departamento de F\'{\i}sica, Instituto Superior T\'ecnico -- IST, Universidade de Lisboa -- UL, Avenida Rovisco Pais 1, 1049-001 Lisboa, Portugal}

\begin{abstract}
The recent gravitational-wave detection GW250114 by the LIGO-Virgo-KAGRA (LVK) Collaboration provides unprecedented opportunities for precision tests of general relativity (GR) through black hole ringdown. 
In this study, we provide the first bounds on theory-agnostic deviations from the Teukolsky equation as described by the beyond-Teukolsky formalism. 
It directly connects deviations in the perturbation equations on the level of the effective potential in the Teukolsky equation with changes in the quasinormal mode (QNM) spectrum. 
We incorporate information on the final mass and spin from a full LVK inspiral-merger-ringdown analysis as parametrized priors in our analysis, reflecting theoretical uncertainties. 
Using publicly available LVK posterior information on agnostic damped sinusoid parameters, we then demonstrate how much beyond-Teukolsky potentials can be constrained. 
The high signal-to-noise ratio (SNR) allows us to avoid the expensive full Bayesian analysis of all parameters and to work directly with a simplified likelihood for the fundamental QNM only. 
This strategy is promising for future events with even higher SNR and allows, in principle, for a quick and simple test of GR without performing the full data analysis procedure, and provides a basis for complementary theory-specific tests. 
We report that current bounds on deviation parameters are in agreement with the Teukolsky equation. 
\end{abstract}

\maketitle

\section{Introduction}

Since the first detection of gravitational waves~\cite{LIGOScientific:2016aoc,LIGOScientific:2018mvr,LIGOScientific:2020ibl,LIGOScientific:2021usb,KAGRA:2021vkt,LIGOScientific:2025slb}, black hole spectroscopy has become one of the central pillars for testing general relativity (GR)~\cite{Detweiler:1980gk,Echeverria:1989hg,Finn:1992wt,Dreyer:2003bv,Berti:2005ys,Carullo:2019flw,Brito:2018rfr,Ma:2022wpv,Isi:2021iql}; see Refs.~\cite{Kokkotas:1999bd,Nollert:1999ji,Berti:2009kk,Konoplya:2011qq,Berti:2018vdi,Franchini:2023eda,Berti:2025hly} for extended reviews on this topic. 
The gravitational waves emitted after the merger of two astrophysical black holes consist of a prompt response, followed by a superposition of characteristic damped oscillations known as quasinormal modes (QNMs), and conclude with a superposition of tails. 
In GR, the spectrum of a Kerr black hole~\cite{Kerr:1963ud} can be computed as a boundary value problem in black hole perturbation theory, through the Teukolsky equation~\cite{Teukolsky:1972my,Teukolsky:1973ha}. 
The spectrum of QNMs is uniquely described by the mass and spin of the BH: measuring more than one QNM allows one to test for consistency, and falsify the no-hair theorem's assumptions~\cite{Israel:1967wq,Hawking:1971vc,Carter:1971zc,Robinson:1975bv}.

The validity of linear perturbation theory to describe the ringdown of a binary black hole merger has been verified in several cases with numerical relativity simulations~\cite{Buonanno:2006ui,Berti:2007fi,Baibhav:2023clw}. 
Although a superposition of QNMs is never a full solution, considerable work has mapped ringdown parameters, such as the final mass and spin~\cite{Buonanno:2007sv,Rezzolla:2007rz,Tichy:2008du,Barausse:2009uz,Healy:2014yta}, as well as the amplitudes and phases of QNMs~\cite{Kamaretsos:2012bs,London:2014cma,Cheung:2023vki,Pacilio:2024tdl,MaganaZertuche:2024ajz,Carullo:2024smg,Mitman:2025hgy,Nobili:2025ydt,Gao:2025zvl} to premerger binary parameters. Current investigations aim at including in the ringdown signal all the non-QNM contributions, such as nonlinearities~\cite{Sberna:2021eui,Cheung:2022rbm,Mitman:2022qdl,May:2024rrg}, tails~\cite{DeAmicis:2024not,DeAmicis:2024eoy,Ma:2024hzq,Rosato:2025rtr} and prompt response~\cite{DeAmicis:2025xuh,Kuntz:2025gdq,Arnaudo:2025uos,DeAmicis:2026wqd,Arnaudo:2026tcy}.

All the efforts of modeling the ringdown as accurately as possible aim at measuring the parameters of the last stage of a binary black hole merger and use them to constrain GR. For a low signal-to-noise ratio (SNR) event, one could either compare the measurement of the final BH parameters recovered from a ringdown-only analysis against those of a full inspiral-merger-ringdown (IMR) one~\cite{LIGOScientific:2016lio,LIGOScientific:2020tif,LIGOScientific:2021sio,LIGOScientific:2026wpt,LIGOScientific:2026oim}; or combine together the deviation sampled over several events, either for specific theories~\cite{Carullo:2021oxn,Maenaut:2024oci,Chung:2025ucb} or generic deviations~\cite{LIGOScientific:2021sio,LIGOScientific:2026wpt,LIGOScientific:2026oim,Zhong:2024pwb}.  For the first time, with the exceptionally loud event GW250114, it was possible to detect two simultaneous modes and go beyond consistency tests of GR~\cite{LIGOScientific:2025rid,LIGOScientific:2025wao}. 
Additional ringdown tests with full IMR waveform models have been applied to GW250114 in Refs.~\cite{Chandra:2025jfc,Grimaldi:2026prn}.

One may wonder what happens by going beyond null tests of GR with the ringdown of GW250114. Perturbation theory applied to rotating black holes beyond GR is a highly nontrivial problem, as it is limited to a handful of theories~\cite{Pierini:2021jxd,Pierini:2022eim,Wagle:2021tam,Srivastava:2021imr,Cano:2023jbk,Cano:2023tmv,Cano:2024ezp,Khoo:2024agm,Blazquez-Salcedo:2024oek,Chung:2024vaf,Chung:2025gyg,Alapati:2025jmd}. In this paper, we pursue a more agnostic approach, by selecting the spectrum of a modified Teukolsky equation. Such an approach was initially developed in the nonrotating case~\cite{Cardoso:2019mqo,McManus:2019ulj,Kimura:2020mrh,Volkel:2022aca,Volkel:2022khh,Franchini:2022axs,Thomopoulos:2025nuf} and then extended to the Teukolsky equation~\cite{Cano:2024jkd,Tang:2025qaq}, based on the perturbative calculations given in~\cite{Li:2022pcy,Hussain:2022ins,Cano:2023tmv}.

In this work, we propose a simple method that combines IMR analyses and those from the ringdown analysis with a pure damped sinusoid template. By obtaining the mass and the spin of the remnant from the IMR posterior, we predict the distribution of the fundamental mode and damping time. We focus on the fundamental mode only, as it is typically the most robustly and accurately inferred QNM, but in principle, one could apply the same process to any mode. Moreover, the high SNR ratio of the fundamental allows us to approximate the posterior distribution of the measurement using a multivariate Gaussian distribution, simplifying the analysis. We then quantify deviations from GR using the beyond-Teukolsky formalism~\cite{Cano:2024jkd} and resample the ringdown frequencies to constrain the modifications, using the mass and spin IMR distributions as informed priors. Since in alternative theories of gravity the relation between inspiral parameters and the final mass and spin would differ from their GR counterparts, we introduce a scaling parameter that allows for agnostic deviation in the priors. We apply our approach to public LVK results for GW250114~\cite{LIGOScientific:2025wao} to constrain theory-agnostic deviations in the beyond-Teukolsky framework.

The rest of this work is organized as follows. 
We outline our theoretical and data-analysis framework in Sec.~\ref{methods}. 
Application and results are reported in Sec.~\ref{applications}, and we conclude in Sec.~\ref{conclusions}.

\section{Methods}\label{methods}

\subsection{Beyond-Teukolsky framework}\label{meth_bteuk}

We employ the beyond-Teukolsky framework as introduced in Ref.~\cite{Cano:2024jkd}. 
It introduces small parametrized deviations from the Teukolsky equation, which governs the curvature perturbations of Kerr black holes in general relativity.  
The beyond-Teukolsky framework is a generalization of the parametrized QNM framework for nonrotating black holes~\cite{Cardoso:2019mqo, McManus:2019ulj}, which has been studied in more detail in Refs.~\cite{Kimura:2020mrh,Volkel:2022aca,Franchini:2022axs,Hirano:2024fgp,Thomopoulos:2025nuf}. 
It has recently been applied to higher-derivative gravity~\cite{Cano:2024ezp} as one of the few explicitly known perturbation equations of rotating black holes beyond GR, and extended to the time domain~\cite{DeSimone:2026waz}. 

In the framework, one modifies the Teukolsky equation 
\begin{equation}\label{eq:mod_teukolsky}
    \frac{1}{\Delta^s} \frac{\text{d}}{\text{d} r}\left[ \Delta^{s+1} \frac{\text{d} R(r)}{\text{d} r}\right] + V(r)R(r) + \delta V(r) R(r) = 0\,.
\end{equation}
The Teukolsky potential is given by
\begin{align}
    V(r) &= 2 \text{i} s \frac{\text{d} K}{\text{d} r} - \lambda_{\ell m} + \frac{1}{\Delta}\left( K^2 -\text{i} s K \frac{\text{d} \Delta}{\text{d} r} \right) \,,
    \\
    \Delta & = r^2 - 2M r +a^2\,, 
    \\
    K &= (r^2+a^2)\omega - a m \,, \\
    \lambda_{\ell m} & \, = B_{\ell m} + a^2\omega^2 -2 a m \omega \,,
\end{align}
where $B_{\ell m}$ is the separation constant. 
The modification $\delta V(r)$ is assumed to be small and can be expressed in powers of $r$
\begin{equation}
\label{eq:deltav-fuc1}
    \delta V(r) =  \frac{1}{\Delta}\sum_{k=-K}^{4} \alpha_{k} \left( \frac{r}{r_+} \right)^k \,.
\end{equation}
This assumes that the parameters $|\zeta_k| \ll 1$, where $\zeta_k = \alpha_k/M^2$, are small complex-valued constants parametrizing deviations, $K>0$ is an arbitrary integer and the location of the outer horizon given by $r_+ = M+\sqrt{M^2-a^2}$.
The QNMs are then connected, at linear order, to the modification via
\begin{align}
    \omega =  \omega^\text{GR} + \frac{1}{M} \sum_{k=-K}^{4} \zeta_k \, d^k_\omega \label{eq:domega}\,, 
\end{align}
where we omitted the explicit $(\ell, m, n)$ dependence and the coefficients $d^k_\omega$ are provided in the repository~\cite{sebastian_volkel_2024_14001739}. In the following equations, we will work with the dimensionless spin $\chi = a/M$.
For more details on the development and application of the framework, as well as the computation of $d^k_\omega$, we refer to Ref.~\cite{Cano:2024jkd}.

\subsection{Likelihood and LVK data}

\begin{figure}
\centering
\includegraphics[width=1.0\linewidth]{"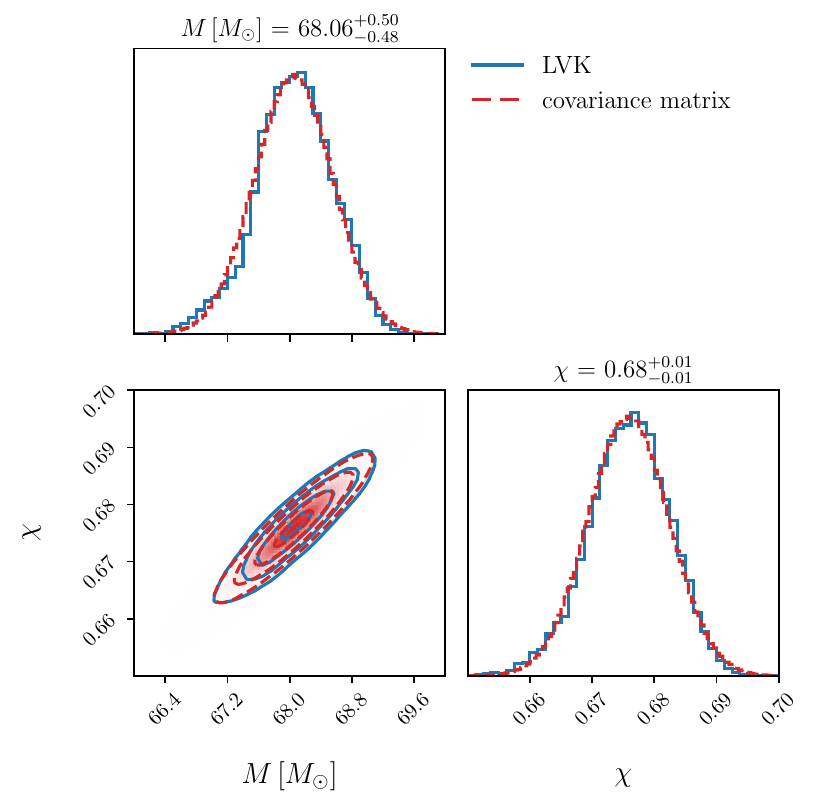"}
\includegraphics[width=1.0\linewidth]{"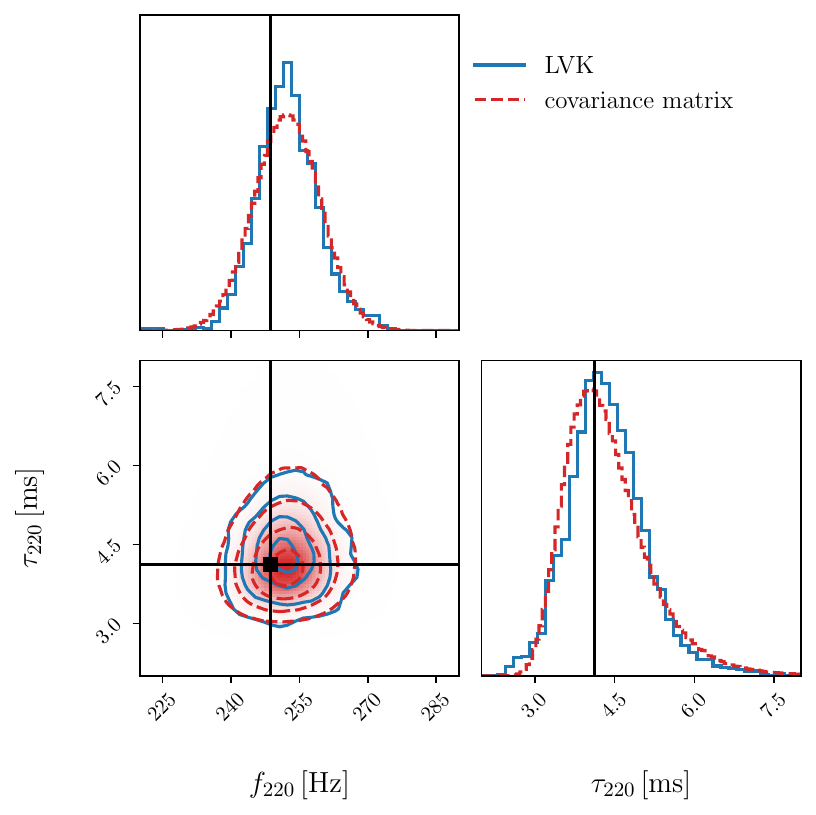"}
\caption{Marginalized posterior distributions from the LVK analysis (blue) 
reported in Ref.~\cite{LIGOScientific:2025wao} along with samples from approximating them with multivariate Gaussians centered at the maximum-likelihood values (red). 
Top: final mass and spin from the NRSur7dq4 analysis.
Bottom: fundamental mode frequency and damping time of the theory-agnostic analysis of two damped sinusoids starting at $t=10\,M$ converted as in Eq.~\eqref{lvk_omega}. 
The full-IMR maximum-likelihood prediction for the fundamental mode is shown by the black crosshairs. 
}
\label{fig:1}
\end{figure}

We approximate the posterior distributions of mass and spin by a multivariate Gaussian
\begin{align}\label{eq:multivariate_gaussian_mass_spin}
p(M,\chi) = \mathcal{N}\left[(M,\chi), \mu, \sigma \right]\,,
\end{align}
where $\mu = (\hat{M}, \hat{\chi}) = (68.1\,M_\odot, 0.68)$ is a vector containing the maximum likelihood value of the mass and spin posteriors, respectively and $\sigma=\sigma_{M\chi}$ is their covariance matrix. 
The values of $\mu$ and $\sigma$ are computed from the publicly available NRSur7dq4 samples~\cite{Varma:2019csw} from the full IMR analysis of GW250114~\cite{LIGOScientific:2025wao}. Analogously, we approximate the posteriors of $f_{220}$ and $\tau_{220}$ given by
\begin{align}\label{lvk_omega}
\omega^\text{data}_{220} = 2 \pi f_{220} - \mathrm{i} \frac{1}{\tau_{220}}\,.
\end{align}
Since we only focus on the fundamental mode, we use posteriors of $\omega^\text{data}_{220}$ from the theory-agnostic analysis of two damped sinusoids starting at $t=10\,M$. The apparent skewness of $\tau_{220}$ is because the LVK posteriors are for $1/\tau_{220}$. 

In Fig.~\ref{fig:1}, we compare the multivariate Gaussian against the LVK posteriors for both mass/spin and frequency/damping time pairs. Both cases show very good agreement, allowing us to treat the two probability distributions analytically.
The agreement is better for mass and spin, as it shows small deviations for the fundamental mode. 
However, we consider it to be within the approximate nature of our work. 

Let us comment on the decision to focus solely on the fundamental mode. In principle, the analysis of GW250114 also confirms the presence of the first overtone $(\ell, m ,n) = (2,2,1)$ and suggests tentative evidence for the $(2,2,2)$ and $(4,4,0)$ modes. The constraints on those modes, though, are much looser, and their distribution is far from well approximated by a multivariate Gaussian. Moreover, by construction, the free parameters $\zeta_k$ of the beyond-Teukolsky formalism are frequency-dependent values and hence would change for each mode number.

For our analysis, we define the likelihood for $\omega_{220}$ as follows
\begin{align}\label{likelihood}
\log \left[ p( \omega^\text{data}_{220} | \theta)  \right] &= -\frac{1}{2} h \sigma_{\omega_{220}}^{-1} h \,,
\end{align}
with 
\begin{align}
h= \omega^\text{model}_{220}(\theta)-\omega^\text{data}_{220}\,,
\end{align}
where $h$ denotes the two-component vector containing the real and imaginary parts, $\theta = [M, \chi, \text{Re}(\zeta_k), \text{Im}(\zeta_k) ]$, and $\sigma_{\omega_{220}}$ is the $2 \times 2$ covariance matrix computed from the agnostic ringdown chains. 
We adopt uniform priors for the real and imaginary part of $\zeta_k$ between $[-8,8]$ and only vary one complex deviation parameter at a time (real and imaginary parts of $\zeta_k$ are varied simultaneously). 
This choice follows standard practice in parametrized tests of GR and avoids severe parameter degeneracies, see e.g. Ref.~\cite{LIGOScientific:2025wao}.

\subsection{Informed priors for mass and spin}

The second part concerns priors for final mass and spin needed to predict the QNM spectrum, even within GR. 
Using only the real and imaginary parts of the fundamental mode, one could obtain $M$ and $\chi$ assuming GR is valid, but this would not leave room to constrain further any deviation parameters (same number of known and unknown parameters). 
However, if deviations from GR are small, one could argue that the final mass and spin are similar to the ones predicted by a full IMR analysis (small ``theoretical errors''). 
This implies that the final mass and spin are not completely unknown, and one can use this assumption to incorporate prior knowledge. 
We define this uncertainty in terms of the definition of mass and spin in two different ways.

\emph{Prior 1---}
The first prior choice is a slightly modified multivariate Gaussian distribution as in Eq.~\eqref{eq:multivariate_gaussian_mass_spin}, but where we substitute $\sigma$ with $\sigma_\lambda = \lambda \sigma$ with $\lambda$ a positive real number.
The multiplication factor $\lambda$ accounts for our ignorance of how much the final mass and spin differ from the GR-based IMR predictions. It quantifies the theoretical error.
If one assumes that the statistical errors on $M$ and $\chi$ are much larger than the theoretical ones, one would set $\lambda \approx 1$, while $\lambda \gg 1$ gives less informative priors and thus relaxes the GR dependency. 

In the extreme limit of setting it to zero, one fixes the final mass and spin to the IMR maximum-likelihood value, constraining only deviations from the fundamental mode. 
To some extent, this is qualitatively similar to how bounds of PN-deviation parameters from the inspiral are inferred; they are typically varied only one at a time~\cite{LIGOScientific:2021sio}. 
The priors on the spin must also satisfy the additional physical requirement of being bounded in $[0,1]$.

\emph{Prior 2---} 
To allow for a second type of prior, we also sample final mass and spin from truncated uniform priors such that mass and spin are in a box centered at the maximum likelihood value described by
\begin{align}
[M, \chi] \in [ (1-\lambda/100)[\hat{M}, \hat{\chi}], (1+\lambda/100)[\hat{M}, \hat{\chi}]]\,,
\end{align}
where $\lambda$ is again parametrizing our ignorance. 
It should be understood as a proxy for the marginalized uncertainty across possible beyond-GR theories. 
In GR, the difference between the sum of the inspiral component masses typically differs by only a few percent from the final mass. 
The prediction of final spin from the inspiral parameters is less trivial and, in general, more complex. 
However, working under the hypothesis that changes with respect to GR predictions are small, and for simplicity, we treat them in the same way. 

For both priors, one would expect wider posterior distributions for $\zeta_k$ with increasing values of $\lambda$.

\section{Application and results}\label{applications}

Using the previously described setup in Sec.~\ref{methods}, we sample the likelihood Eq.~\eqref{likelihood} with the two types of theory-agnostic priors using the \texttt{EMCEE} sampler~\cite{Foreman-Mackey:2012any} based on the affine-invariant ensemble sampler~\cite{2010CAMCS...5...65G}. 
In each analysis, we initialized the sampler using 100 walkers, 10000 samples, a burn-in of 500, and thinned the resulting chains by 10.
 
\begin{figure*}
\centering
\includegraphics[width=1.0\linewidth]{"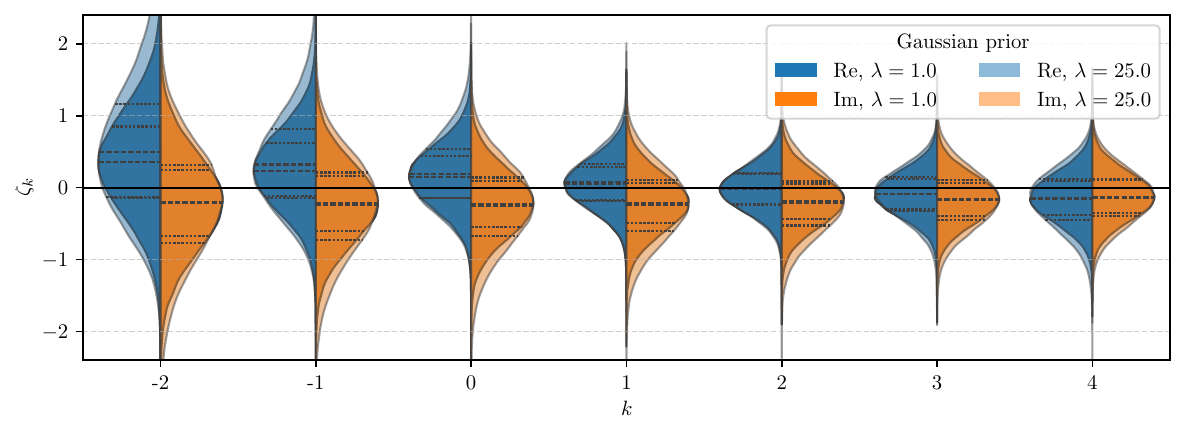"}
\caption{Optimistic ($\lambda=1$) and pessimistic ($\lambda=25$) bounds for the deviation parameters $\zeta_k$ after incorporating the LVK ringdown analysis of GW250114. 
The left and right violins show the posterior distributions of the real and imaginary parts of $\zeta_k$ for different $k$. 
Horizontal lines show quantiles of [0.25, 0.5, 0.75].}
\label{gaussian_violins}
\end{figure*}

In Fig.~\ref{gaussian_violins}, we first show the results for the Gaussian prior. 
Each violin represents the marginalized posterior distributions for the real and imaginary parts of various $\zeta_k$ and two different choices for $\lambda$. 
Note that the uniform priors for real and imaginary part of $\zeta$ are between $[-8, 8]$, and are indeed fairly uninformative compared to the posteriors. 
The bounds for $\lambda=1$ can be understood as ``optimistic'' bounds, since they completely ignore the theoretical uncertainties in final mass and spin, and underestimate the statistical errors since they use the full IMR information.
Consequently, the bounds for $\lambda=25$ can be called ``pessimistic bounds'', because they accommodate theoretical errors for final mass and spin that are $\sqrt{\lambda}=5$ times larger than the statistical ones. 
We verified that values of $\lambda \ll 1 $ agree with the trivial Gaussian error propagation that only maps the errors of $\sigma_{\omega_{220}}$ to a fixed $\sigma_{\zeta_{k}}$ by fixing mass and spin to the maximum likelihood values.

Most importantly, we report that all complex deviation parameters $\zeta_k$ are in good agreement with GR ($\zeta_k=0$); even when considering the more narrow optimistic bounds. 
The scaling behavior for different $\zeta_k$, indicating that bounds are more stringent for larger $k$, is qualitatively expected. 
Different radial powers indexed by $k$ (for the same dimensionless number $\zeta_k$) modify the fundamental mode strongly, thus resulting in narrower posteriors.

\begin{figure*}
\centering
\includegraphics[width=1.0\linewidth]{"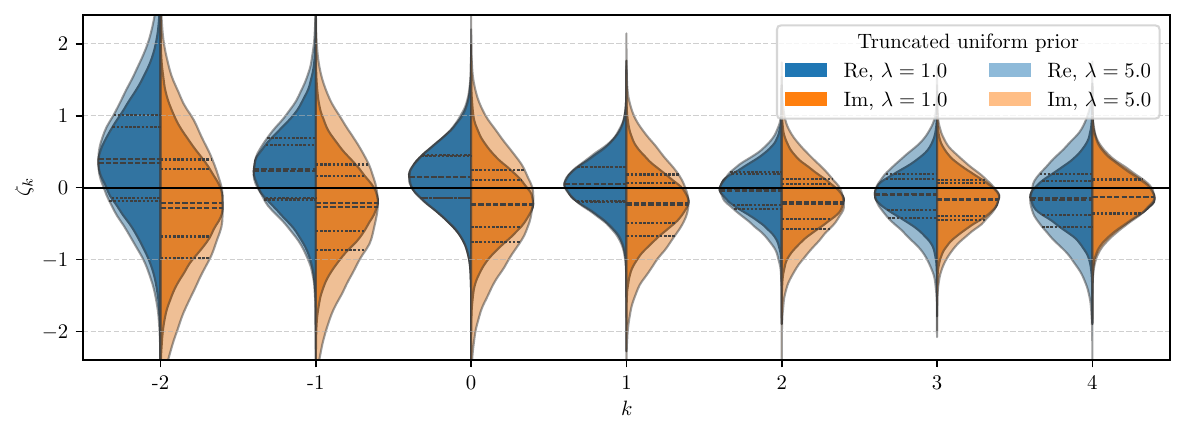"}
\caption{Bounds on $\zeta_k$ from truncated uniform priors, similar structure as in Fig.~\ref{gaussian_violins}. 
}
\label{box_violins}
\end{figure*}

Next, we discuss the results for the truncated uniform prior, for which the corresponding violin plots are presented in Fig.~\ref{box_violins}. 
Overall, most trends are similar to the ones with a Gaussian prior, especially for $\lambda=1$. 
Differences mainly arise for $\lambda=5$, for which some of the real or imaginary parts can become more impacted, like the real part of $\zeta_4$. In contrast, some others are very mildly impacted, like the imaginary part of $\zeta_4$. 

Since the beyond-Teukolsky framework~\cite{Cano:2024jkd} provides us with linearized corrections to the QNMs for a given $\zeta_k$, let us comment on the validity of our analysis. 
In Fig.~2 of Ref.~\cite{Cano:2024jkd}, it is shown that linear corrections are about 1\,\% accurate for the here considered ranges of $\zeta_k$.
Since higher-order corrections in $\zeta_k$ are currently not provided, a related extension of our analysis could be the subject of future work. 
Moreover, note that the constraints presented here correspond to scales of a few tens of km. Taking the square root of the one-standard-deviation uncertainty of the marginalized real or imaginary part of $\zeta_k \hat{M}^2$ yields values in the range of approximately $60$--$100\,\mathrm{km}$, with consistently stronger constraints for larger $k$.
Our analysis is complementary to the agnostic analysis of parametrized spin expansion coefficients (ParSpec)~\cite{Maselli:2019mjd,Carullo:2021dui}, applied to GW250114 in Ref.~\cite{Chen:2026ack}. The reported bound of the characteristic length of deviation from GR is around $80\,\mathrm{km}$, which is consistent with our reported bounds. 

Finally, we note that the here presented analysis has conceptually been inspired by earlier works~\cite{Volkel:2020daa,Volkel:2022aca,Volkel:2022khh}, which used high-precision QNM mock data, simplifying assumptions about measurement errors, and focused on nonrotating black holes to bound deviation parameters of the effective potential or parameters of the Rezzolla-Zhidenko metric~\cite{Rezzolla:2014mua}. 
In Ref.~\cite{Volkel:2022khh}, violin plots for real-valued deviation parameters in the spherically symmetric (nonrotating) case have been studied for GR injections (Fig.~2) and non-GR injections (Fig.~4). 
In case of a true GR violation (or systematic effects in the LVK analysis, e.g., biased final mass and spin or start of the ringdown), one would expect that the here presented bounds for $\zeta_k$ (one at a time), would, in general, show beyond-GR deviations for all $\zeta_k$, even if only one or a few of them are actually nonzero. 
Varying many $\zeta_k$ simultaneously may seem to yield uninformative bounds (one- or two-dimensional marginals), but in the nonrotating case, it has been shown that such posteriors are indeed informative when mapped to local properties of the effective potential around its maximum; as expected from WKB theory (Fig.~3 and Fig.~5 in Ref.~\cite{Volkel:2022khh}).  
Although the higher-order WKB method can also be extended to the more complicated case of the Teukolsky equation~\cite{Seidel:1989bp,Kokkotas:1991vz} and the beyond-Teukolsky framework~\cite{Tang:2025qaq}, a similar mapping is left for future work.

\section{Conclusions}\label{conclusions}

Precision measurements of gravitational waves from binary black hole mergers have become a standard tool for probing strong gravity. 
In this work, we motivate and outline a simple approach suitable for black hole spectroscopy with high-SNR events, such as GW250114~\cite{LIGOScientific:2025wao}. 
Our analysis combines IMR results for the final mass and spin of a black hole with those of theory-agnostic damped sinusoids applied to the ringdown. 

In general, the final black hole mass and spin differ from those predicted from the inspiral parameters when considering theories beyond GR. 
Since this difference is theory-dependent and typically unknown, we introduced two parametrizations to incorporate prior knowledge from GR. 
We then showed how the theory-agnostic posterior of the fundamental QNM can be used to constrain the complex-valued deviation parameters of the beyond-Teukolsky framework~\cite{Cano:2024jkd}. 
The latter explicitly connects deviations in the QNM spectrum to possible deviations of the effective potential in the Teukolsky equation~\cite{Teukolsky:1972my,Teukolsky:1973ha}; the standard model for black hole perturbation theory in GR. 
The basic steps of the analysis are not limited to the beyond-Teukolsky framework and could also be repeated for any application where the fundamental QNM is a function of mass, spin, and additional parameters. 

As a concrete example, we applied our framework to the publicly available posteriors of the LVK Collaboration analysis for GW250114. 
In particular, we used the NRSur7dq4 posteriors for final mass and final spin, and the fundamental mode obtained from the two-damped-sinusoid analysis. 
Our analysis demonstrates that all beyond-Teukolsky deviation parameters, when varied one at a time, are in agreement with GR. 
While this should be expected from other tests of GR conducted by the LVK Collaboration~\cite{LIGOScientific:2025wao,LIGOScientific:2025rid}, our work provides the first bounds on theory-agnostic deviations from the Teukolsky equation itself. 
It would be interesting to assess whether future high-SNR detections and next-generation detectors could enable this framework to probe beyond-Teukolsky deviations with substantially improved precision, and potentially distinguish between theory-specific signatures.

Future extensions of this work could make the here-presented modeling of systematic uncertainties in the final mass and spin obtained from a full IMR analysis (or possible inspiral predictions only) more quantitative, given the knowledge of how strong typical deviations from GR can be in theory-specific cases. 
Moreover, in this work, we only focused on the fundamental mode and did not directly include overtones. 
This is because overtones are more difficult to extract, and because deviation parameters in the beyond-Teukolsky framework do, in general, depend on each QNM frequency. 
An application to theory-specific cases may evade this limitation, as the agnostic deviation parameters would not be independent but controlled by a single coupling parameter. 
Future work should also propagate sky-localization uncertainty in the ringdown inference~\cite{Dey:2026btx}.

\acknowledgments
S.\,H.\,V. thanks Ciro De Simone, Arnab Dhani, and Pratik Wagle for useful discussions and valuable feedback on the manuscript.  
S.\,H.\,V. acknowledges funding from the Deutsche Forschungsgemeinschaft (DFG): Project No. 386119226.
N.F.~acknowledges funding from the FCT grant agreement 2023.06263.CEECIND/CP2830/CT0004 and support to the Center for Astrophysics and Gravitation (CENTRA/IST/ULisboa) through FCT grant No.~UID/PRR/00099/2025 and grant No.~UID/00099/2025.

\bibliography{literature}

\end{document}